\documentclass[pre,reprint,a4paper,superscriptaddress]{revtex4-2}
\usepackage{amsmath}
\usepackage{graphicx}

\usepackage{bm}
\usepackage{float}
\usepackage{hyperref}

\DeclareMathOperator{\re}{Re}
\DeclareMathOperator{\im}{Im}

\begin{document}
\title{Dynamics of interacting cavity solitons}
\author{Amir Leshem} 
\affiliation{The Racah Institute of Physics, The Hebrew University of Jerusalem, Jerusalem, Israel 9190401}\affiliation{Pre-Academic Preparatory Studies, Jerusalem Multidisciplinary College, Hanevi’im St 37, Jerusalem 9101001
}
\author{Sanzida Akter}
\author{Logan Courtright}
\author{Pradyoth Shandilya}
\author{Curtis R. Menyuk}
\affiliation{Computer Science and Electrical Engineering Department, University of Maryland at Baltimore County, 1000 Hilltop Circle, Baltimore, Maryland 21250, USA}
\author{Omri Gat}
\email{omrigat@mail.huji.ac.il} 
\affiliation{The Racah Institute of Physics, The Hebrew University of Jerusalem, Jerusalem, Israel 9190401}

\begin{abstract}
We derive the equations governing the motion of Kerr solitons in pair waveforms. Recent experiments in microresonators have studied a variety of interaction effects in multisoliton waveforms, including collisions and formation of soliton molecules and crystals. Here we analyze the effective interaction that arises from the coupling of soliton-tail overlap nonlinearity with global soliton variables associated with the breaking of translation symmetry. The interaction is either purely repulsive, or alternates between attraction and repulsion, according to whether the decay of soliton tails is monotone or oscillatory. In the latter case, stable fixed points of the effective dynamical system signify stable soliton molecule configuration, but the exponential weakening of the interaction with increasing inter-soliton separation may prevent the molecule from forming in experimentally accessible time scales. Our theory becomes asymptotically exact in the large-separation limit, and we verify the theoretical calculations using soliton trajectories extracted from direct numerical solutions of the wave equation.

\end{abstract}
\maketitle
\section{Introduction}
Solitons are a wave analog of composite particles since they are emergent structures that when weakly forced, respond through the evolution of global degrees of freedom associated with broken symmetries while maintaining their shape. Optical solitons in pumped Kerr cavities have been intensively studied since it was demonstrated that they can be used to generate optical frequency combs by Herr et al.\ \cite{herrTemporalSolitonsOptical2014}. While single-soliton waveforms are the simplest comb sources, there has been great interest in multisoliton waveforms, with experiments demonstrating complex soliton dynamics \cite{guoUniversalDynamicsDeterministic2017,yiImagingSolitonDynamics2018}, formation of soliton molecules \cite{wangUniversalMechanismBinding2017,wengHeteronuclearSolitonMolecules2020} and soliton crystals \cite{coleSolitonCrystalsKerr2017,karpovDynamicsSolitonCrystals2019,hePerfectSolitonCrystals2020,luSynthesizedSolitonCrystals2021,huSpatiotemporalBreatherDynamics2024,shandilyaAllopticalAzimuthalTrapping2025}. 

Even though cavity solitons are observed to propagate as discrete, particle-like objects, their dynamics has so far been almost exclusively studied on the basis of wave equations that do not directly describe the motion of the soliton degrees of freedom. Here we derive the effective equations of motion for two interacting Kerr solitons. The nonlinearities of the wave equation generate overlap terms that couple to the global soliton degrees of freedom through the zero modes in the spectrum of the soliton stability operator that are associated with the broken translation symmetry.

This approach was originally applied to solitons in perturbed integrable systems \cite{karpmanSolitonSystemSubject1979,karpmanPerturbationalApproachTwosoliton1981}, and then generalized to kinks and pulses in reaction-diffusion and other pattern forming system \cite{gorshkovInteractionsSolitonsNonintegrable1981, kawasakiKinkDynamicsOnedimensional1982, carrMetastablePatternsSolutions1989,aransonStableParticlelikeSolutions1990, eiEquationMotionInteracting1994,  shin-ichiroeiMotionWeaklyInteracting2002, vladimirovStableBoundStates2001, turaevChaoticBoundState2007}. It is also applicable to the interaction of solitons with an external force \cite{mizrahiUniversalDynamicsSpatiotemporal2023,mizrahiSolitonSynchronizationMicroresonators2024}. An alternative approach based on a variational principle \cite{malomedBoundSolitonsNonlinear1991,akhmedievMultisolitonSolutionsComplex1997}, has also been used to study soliton interactions, but since it involves an uncontrolled approximation, the accuracy of the equations of motion derived using the variational method is limited.

In this work we study the evolution of two-soliton waveforms of the basic  Lugiato-Lefever model of Kerr resonators \cite{chemboSpatiotemporalLugiatoLefeverFormalism2013}. Since the global phase of the cavity solitons is locked to the pump, translation is the only continuous symmetry, and the associated degrees of freedom are the soliton positions. Since the model also has a reflection symmetry, the phase space reduces to the single degree-of-freedom of soliton separation, and the effective dynamics becomes simple. 

In spite of this simplicity and the interest in applications to Kerr resonator experiments, there have been few studies of soliton interactions in this system. The investigations of \cite{barashenkovExistenceStabilityChart1996} and \cite{parra-rivasInteractionSolitonsFormation2017} focused on solitons with moderate detuning that have limited experimental relevance, and used the variational method. A recent study, reported in \cite{vladimirovDissipativeSolitonInteraction2021}, focused on the effects of high-order dispersion.

Here we derive effective equations of motion for the soliton positions that are asymptotically exact for well-separated solitons. There are only two possible types of asymptotic interaction between soliton pairs where translation is the only continuous symmetry, depending on whether the decay of the soliton tail is monotone or oscillatory. Both interaction types occur with red-detuned cavity solitons for appropriate choices of system parameters, although oscillatory tails are much more common \cite{parra-rivasOriginBifurcationStructure2021}; we find that when the soliton tails are monotone, the soliton interaction is purely repulsive and decreases exponentially, while the relative velocity oscillates sinusoidally within an exponentially decreasing envelope when the soliton tails are oscillatory. In the latter case, the interaction alternates between intervals of repulsion and attraction of equal length, separated by interlaced lattices of stable and unstable fixed points. Even though each stable fixed point corresponds to a stable two-soliton molecule, the strength of interaction between highly red-detuned solitons decreases rapidly as their separation increases, which implies that only tightly bound molecules can form within experimentally realizable time intervals.



We present the main elements of our theory in section \ref{sec:drift}, where we introduce the model and review the properties of the cavity solitons and their dynamical spectrum. We identify the overlap parameter as the small parameter that governs the asymptotic approximation, and calculate the overlap terms; we then project the overlap terms on the adjoint zero mode of translation to derive the effective equations of motion. In section \ref{sec:tail}, we use the soliton tail asymptotics to simplify the projection integrals and derive explicit equations of motion for the soliton separation, whose parameters are determined only by the single-soliton waveform. The drift velocities derived from the overlap-projection theory are then compared with those extracted from direct numerical simulations of the wave equation, showing the expected agreement in the large-separation regime. Finally, section \ref{sec:conclusions} presents the conclusions and implications of this work.



\section{The soliton drift velocity}\label{sec:drift}
\subsection{Cavity solitons and their linear stability}
We model the evolution of the cavity electromagnetic field with the basic Lugiato-Lefever model, using the normalization of \cite{mizrahiSolitonSynchronizationMicroresonators2024}
\begin{equation}\label{eq:lle}\begin{split}
\frac{\partial\psi}{\partial t}&=N[\psi]\ ,\\ N[\psi]&\equiv-\Bigl(\delta+\frac{i}{2}\Bigr)\psi+ \frac{i}{2}\frac{\partial^2\psi}{\partial x^2}+i|\psi|^2\psi-\frac{ih}{2}\ ,
\end{split}\end{equation}
where $\delta$ is the normalized loss coefficient, and $h=h_r+i(4/\pi)\delta$ is the amplitude of the continuous-wave pump ($\delta$ and $h_r$ are real and positive). For sufficiently small $\delta$ and $h_r$ equation \eqref{eq:lle} has a stable soliton solution $\psi_s(x)$, $N[\psi_s]=0$, that is unique up to translations. The cavity solitons ride on a pedestal
\begin{equation}
\lim_{x\to\pm\infty}\psi_s=\psi_c\ ,
\end{equation}
where $\psi_c$ is a stable continuous-wave solution of Eq.\ \eqref{eq:lle}. Parity symmetry implies that $\psi_s$ is an even function $\psi_s(-x)=\psi_s(x)$.

Since the tails of solitons are weak, $\tilde\psi_s(x)$ is governed by a system of linear equations with constant coefficients for $|x|\gg1$, from which it follows \cite{parra-rivasOriginBifurcationStructure2021} that for such $x$
\begin{equation}\label{eq:vmr}
\tilde\psi_s(x)\mathop{\sim} a e^{-\sigma|x|}\ ,
\end{equation}
or
\begin{equation}\label{eq:vmi}
\tilde\psi_s(x)\sim a_+ e^{-\sigma|x|}+a_- e^{-\sigma^*|x|}\ ,
\end{equation}
depending on whether
\begin{equation}
\sigma=\sqrt{1-4|\psi_c|^2-2\sqrt{|\psi_c|^4-\delta^2}}
\end{equation}
is real or complex (respectively.) In either case the outer square root sign is chosen so that $\re\sigma>0$, and when $\sigma$ is complex, the inner square root sign is chosen such that $\im\sigma>0$. When $\sigma$ is real the tails of $|\psi_s|$ decay monotonically, while they decay with oscillations when $\sigma$ is complex. An analysis of the stable soliton parameter space shows \cite{parra-rivasOriginBifurcationStructure2021} that there are highly red-detuned solitons with both types of tails, although the oscillatory tails are more common and are the ones mostly seen in experiments.

The linear stability of the cavity soliton solutions is convenient to study in a doubled phase space of vector values functions $\Psi=(\psi,\psi^*)$ that obeys
\begin{equation}\label{eq:lled}\begin{split}
\frac{\partial\Psi}{\partial t}&=\mathcal{N}[\Psi]\ ,\\\mathcal{N}[\Psi]&\equiv-\Bigl(\delta+\frac{I}{2}\Bigr)\Psi+ \frac{I}{2}\frac{\partial^2\Psi}{\partial x^2}+I\Psi^2\Psi^*-\frac{I}{2}(h,h^*);\end{split}
\end{equation}
here multiplication of vectors is defined component-wise, $I=\bigl(\begin{smallmatrix}i&0\\0&-i\end{smallmatrix}\bigr)$, and $\Psi^*=\bigl(\begin{smallmatrix}0&1\\1&0\end{smallmatrix}\bigr)\Psi=(\psi^*,\psi)$. The complexified phase space includes vectors that are not in the subspace where $\psi^*$ is the complex conjugate of $\psi$, but if $\Psi$ is in this subspace initially, then so are $\Psi_t$ for later $t$.

The linear operator that generates the evolution of small perturbations around a single-soliton waveform is
\begin{equation}\label{eq:lles}\begin{split}
\frac{\partial\Phi}{\partial t}&=\mathcal{L}\Phi+O(\Phi^2)\ ,\\\mathcal{L}&= \begin{pmatrix}
A & B\\ B^* & A*
\end{pmatrix}\ ,\end{split}
\end{equation}
\begin{equation}\label{eq:lles}\frac{\partial\Phi}{\partial t}=\mathcal{L}\Phi+O(\Phi^2)\ ,\qquad\mathcal{L}= \begin{pmatrix}
A & B\\ B^* & A^*
\end{pmatrix}\ ,
\end{equation}
with
\begin{equation}\label{eq:lleAB}
A=
-\delta+\frac{i}{2}\Bigl(-1+\frac{\partial^2}{\partial x^2}\Bigr)+2i|\psi_s|^2\ ,\quad B= i \psi_s^2\ .\end{equation}

The translation symmetry of Eq.\ \eqref{eq:lled} implies that $\Psi_s(x-\xi)$ is a stable stationary solution of the equation for any shift $\xi$. It then follows that $\Psi_\xi=-d\Psi_s/dx$ is a zero mode of $\mathcal{L}$, so that $\mathcal{L}\Psi_\xi=0$. The assumption that the soliton is dynamically stable implies that the zero eigenvalue is simple, and that the rest of the spectrum of $\mathcal{L}$ is in the left half of the complex plane.

\subsection{The evolution of two-soliton waveforms}
Since Eq.\ \eqref{eq:lled} is nonlinear, superpositions of cavity solitons are not steady-state solutions, in general. Nevertheless, since the solitons are localized, superpositions of \emph{well-separated} solitons approach a quasi-steady state with slightly deformed soliton waveforms, whose evolution time scale is much longer than the typical dynamical time scales. 

Since the single-soliton waveforms are localized, and the nonlinearity in the wave equation \eqref{eq:lle} is local, we expect that the interaction between well-separated solitons is weak. Recalling that the cavity solitons lie on an extended pedestal, we examine the dynamics of a waveform
\begin{equation}\label{eq:2sol}
\Psi(x,t)=\tilde\Psi_s\bigl(x-\xi_1(t) \bigr)+\tilde\Psi_s \bigl(x-\xi_2(t) \bigr)+\Psi_c+\Phi(x,t)\ ,
\end{equation}
with a soliton separation $|\xi_2-\xi_1|$ that is significantly larger than the width of a soliton; here $\tilde\Psi_s=(\tilde\psi_s, \tilde\psi_s^*)$ and $\Psi_c$ are (respectively) the subtracted soliton and pedestal waveforms as above, and $\Phi$ expresses the deformation of the solitons induced by the interaction. The soliton-tail overlap is then $O(\varepsilon)$, $\varepsilon=e^{-\sigma_r|\xi_2-\xi_1|}$, and consequently the time scale of change of $\xi_1$, $\xi_2$, and $\Phi$ is $\sim\varepsilon^{-1}$; furthermore, $\Phi$ is small, $|\Phi|\sim\varepsilon$, and localized near the two pulses.

The decomposition \eqref{eq:2sol} of the two-soliton waveform is non-unique, because it is possible to shift slightly the soliton centers $\xi_{1,2}$ while absorbing the difference in the residual $\Phi$. We remove this arbitrariness by imposing the auxiliary conditions
\begin{equation}
\langle\bar\Psi_{\xi1},\Phi\rangle=\langle\bar\Psi_{\xi2},\Phi\rangle=0\ ,
\end{equation}
where the inner product is defined by
\begin{equation}
\langle(\psi_1,\psi_2),(\phi_1,\phi_2)\rangle=\frac{1}{2}\int_{-\infty}^\infty dx(\psi_1^*\phi_1+\psi_2^*\phi_2)\ ,
\end{equation}
and $\bar\Psi_{\xi1,\xi2}(x)=\bar\Psi_{\xi}(x-\xi_{1,2})$ are shifts of the normalized zero mode of the adjoint of the single-soliton stability operator
\begin{equation}
\mathcal{L}^\dag\bar\Psi_\xi=0\ ,\qquad \langle\bar\Psi_{\xi},\Psi_\xi\rangle=1\ ;
\end{equation}
since zero is a real eigenvalue, the components of $(\bar\psi_\xi,\bar\psi_\xi^*)$ of $\bar\Psi_\xi$ can be chosen to be complex conjugates. 

The equations of motion for the soliton centers now follow by substituting the ansatz~\eqref{eq:2sol} in the wave equation \eqref{eq:lled} and projecting the result on $\bar\Psi_{\xi1}$ and $\bar\Psi_{\xi2}$. Since the soliton velocities are of $O(\varepsilon)$, terms of higher order in $\varepsilon$ will be neglected. We start with the inner product of $\bar\Psi_{\xi1}$ with the left-hand side of Eq.\ \eqref{eq:lled} which yields
\begin{equation}\label{eq:prolhs}
\frac{d\xi_1}{dt}+\langle\bar\Psi_{\xi1},\Psi_{\xi2}\rangle\frac{d\xi_2}{dt}+\Bigl\langle\bar\Psi_{\xi1},\frac{\partial\Phi}{\partial t}\Bigr\rangle\ .
\end{equation}
Since the solitons are well-separated and, like the solitons, the translational zero modes are exponentially localized near the soliton centers, $\langle\bar\Psi_{\xi1},\Psi_{\xi2}\rangle=O(\varepsilon)$; since ${d\xi_2}/{dt}$ is also of $O(\varepsilon)$, the middle term in \eqref{eq:prolhs} is negligible. Since the residual $\Phi$ is both small and slowly changing, ${\partial\Phi}/{\partial t}=O(\varepsilon^2)$, so that the last term in \eqref{eq:prolhs} is negligible as well, implying that
\begin{equation}\label{eq:predxidt}
\frac{d\xi_1}{dt}=\langle\bar\Psi_{\xi1},\mathcal{N}[\tilde\Psi_{s1}+\tilde\Psi_{s2}+\Psi_c+\Phi]\rangle\ ,
\end{equation}
using the shorthand $\Psi_{s1,2}$ ($\tilde\Psi_{s1,2}$) for the (subtracted) soliton centered at $\xi_{1,2}$ (respectively.) In the theory of pulse interactions, Eq.\ \eqref{eq:predxidt} is often viewed as an integrability condition \cite{aransonStableParticlelikeSolutions1990}.

Expanding the nonlinear term of $\mathcal{N}$ in \eqref{eq:predxidt} yields a large number of terms, most of which either cancel or are of $O(\varepsilon^2)$. These terms can be categorized into a few classes as follows
\begin{enumerate}
\item $O(1)$ terms $\Psi_{s1}^2\Psi_{s1}^*$, $\Psi_{s2}^2\Psi_{s2}^*$, and $\Psi_{c}^2\Psi_{c}^*$, that cancel in combination with other $O(1)$ terms from $\mathcal{N}[\Psi]$ because $\mathcal{N}[\Psi_s]=\mathcal{N}[\Psi_c]=0$.
\item Terms that are linear in $\Phi$ and involve no factors of $\tilde\Psi_{s2}$ combine with other terms of $\mathcal{N}[\Psi]$ that are linear in $\Phi$ to form $\mathcal{L}_1\Phi$ where $\mathcal{L}_1$ is the linearization of $\mathcal{N}$ at $\Psi_{s1}$. These terms are of order $\varepsilon$ but cancel in \eqref{eq:predxidt} because $\langle\bar\Psi_{\xi1},\mathcal{L}_1\Phi\rangle=\langle\mathcal{L}_1^\dag\bar\Psi_{\xi1},\Phi\rangle=0$.
\item Terms that are linear in $\Phi$ and involve factors of $\tilde\Psi_{s2}$ are also of $O(\varepsilon)$ but are suppressed by an $O(\varepsilon)$ factor by the inner product with $\bar\Psi_{\xi1}$ and are therefore negligible.
\item Terms that are quadratic or cubic in $\Phi$ are of $O(\varepsilon^2)$ and therefore also negligible.
\end{enumerate}

Hence, the only nonnegligible terms that remain in the expansion of $\mathcal{N}$ in the right-hand side of \eqref{eq:predxidt} are \emph{cross-terms} that involve at least one factor each of $\tilde\Psi_{s1}$ and $\tilde\Psi_{s2}$ and are independent of $\Phi$. Of these, only the ones which are linear in $\tilde\Psi_{s2}$ are of leading order in $\varepsilon$ after taking the inner product with $\bar\Psi_{\xi1}$. Collecting these terms yields the equation of motion
\begin{multline}\label{eq:dxi1dt}
\frac{d\xi_1}{dt}=\bigl\langle\bar\Psi_{\xi1},-I\bigl(2(\tilde\Psi_{s1}\tilde\Psi_{s1}^*+\tilde\Psi_{s1}\tilde\Psi_{c}^*+\tilde\Psi_{c}\tilde\Psi_{s1}^*)\Psi_{s2}\\+(\tilde\Psi_{s1}^2+2\tilde\Psi_{s1}\tilde\Psi_{c})\Psi_{s2}^*\bigr)\bigr\rangle\\=\im\int_{-\infty}^\infty dx\,\bar\psi_{\xi1}^*\bigl(2(|\tilde\psi_{s1}|^2+2\re(\psi_c^*\tilde\psi_{s1}))\tilde\psi_{s2}\\+(\tilde\psi_{s1}^2+2\tilde\psi_{s1}\psi_c)\tilde\psi_{s2}^*\bigr)\ .
\end{multline}
The equation of motion for the second soliton is obtained by exchanging indices $1\leftrightarrow2$ everywhere in Eq.\ \eqref{eq:dxi1dt}.

\section{Tail asymptotics and the soliton pair dynamics}\label{sec:tail}
The equation of motion \eqref{eq:dxi1dt} and its analog with the roles of 1 and 2 exchanged are the solution to the pulse interaction problem in the sense that they allow one to calculate the soliton velocities using only single-soliton data. However, the result is not explicit because the velocities are obtained as an integral that has to be calculated numerically for each value of the soliton separation.

Fortunately, for the case of well-separated solitons that we are considering, we can simplify the integral in \eqref{eq:dxi1dt} by observing that it is dominated by $x$ values near $\xi_1$, and that for such $x$, $\psi_2$ is well-approximated by its known tail asymptotics.
 Before proceeding to make use of this simplification, we note that reflection and translation symmetries imply that $d\xi_2/dt=-d\xi_1/dt$, and that the velocities depend only on the separation between the solitons, so that without loss of generality we assume $\xi_1=\xi=-\xi_2$ and write the equation of motion for $\xi$
\begin{multline}\label{eq:dxidt}
\frac{d\xi}{dt}=\im\int_{-\infty}^\infty dx\,\bar\psi_{\xi}(x)^*\\\times\bigl(2(|\tilde\psi_s(x)|^2+2\re(\psi_c^*\tilde\psi_s(x)))\tilde\psi_s(x+2\xi)\\+(\tilde\psi_s(x)^2+2\tilde\psi_s(x)\psi_c)\tilde\psi_s(x+2\xi)^*\bigr)\ .
\end{multline}
The preceding argument now shows that we may approximate $\tilde\psi_s(x+2\xi)$ by its tail asymptotics in this integral.
We next consider separately the qualitatively different cases of solitons with monotone and oscillatory tails.

\subsection{Monotone tails}
When the system parameters are such that the soliton tail asymptotics are given by \eqref{eq:vmr} with a real $\sigma$, Eq.\ \eqref{eq:dxidt} becomes
\begin{equation}\label{eq:monotone}
\frac{d\xi}{dt}=b e^{-2\sigma\xi}\ ,
\end{equation}
with
\begin{multline}\label{eq:b}
b=\im\int_{-\infty}^\infty dx\,\bar\psi_{\xi}(x)^*\Bigl(2a\bigl(|\tilde\psi_s(x)|^2+2\re(\psi_c^*\tilde\psi_s(x))\bigr)\\+a^*(\tilde\psi_s(x)^2+2\tilde\psi_s(x)\psi_c)\Bigr)e^{-\sigma x}\ .
\end{multline}
The soliton drift velocity in the equation of motion \eqref{eq:monotone} does not change sign, which implies that there are no stationary states in the asymptotic regime. Moreover, in all the cases that we studied, we found that $b$ is positive, so that the soliton interaction is purely repulsive. The solution of \eqref{eq:monotone} can be expressed as
\begin{equation}
\xi(t)=\frac{\log(2\sigma b t)}{2\sigma}\ ;
\end{equation}
the exponential fall-off of the drift speed with increasing separation implies a logarithmically slow growth of the soliton separation.

\begin{figure}
\includegraphics[width=7.5cm]{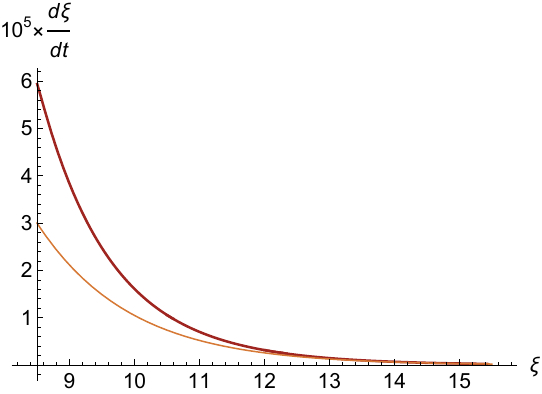}\\[5mm]
\includegraphics[width=7.5cm]{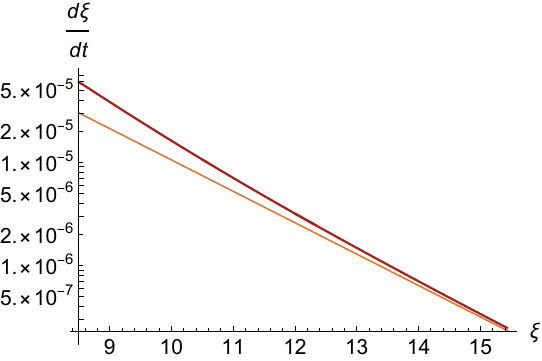}
\caption{\label{fig:monotone} Kerr soliton drift velocity $d\xi/dt$ in a waveform with two solitons centered at $\xi$ and $-\xi$. The normalized cavity parameters are $\delta=0.185$, $h_r=0.209$, such that the solitons have exponentially decaying tails, giving rise to a purely repulsive interaction. The thin light-colored curve shows the theoretically predicted velocity, and the thick dark-colored curves show velocities calculated from soliton trajectories that we obtained from numerical solutions of the wave equation \eqref{eq:lle}, displayed on linear and logarithmic scales in the top and bottom panels (respectively). As expected, the curves converge in the large-separation limit.}
\end{figure}

The effective drift velocity of a monotone-tail soliton pair \eqref{eq:monotone} is compared with that extracted from direct simulations in figure \ref{fig:monotone}. We note that the terms appearing between the large parentheses in the integral in \eqref{eq:b} are all even functions of $x$, while $\bar\psi_\xi$ is odd, so that the integration produces a strong cancellation that degrades the accuracy of the approximations leading to \eqref{eq:monotone}; as a consequence, the effective interaction theory becomes valid only for separations significantly larger than the soliton width.

\subsection{Oscillatory tails}
When the soliton tail asymptotics are given by \eqref{eq:vmi} with a complex $\sigma=\sigma_r+i\sigma_i$, the equation of motion \eqref{eq:dxidt} becomes
\begin{equation}\label{eq:osce}
\frac{d\xi}{dt}= e^{-2\sigma_r\xi}\im\bigl(b_+e^{-2\sigma_i\xi}+b_-e^{2\sigma_i\xi}\bigr)\ ,
\end{equation}
with
\begin{multline}\label{eq:bplus}
b_+=\int_{-\infty}^\infty dx\,\bar\psi_{\xi}(x)^*\Bigl(2a_+\bigl(|\tilde\psi_s(x)|^2+2\re(\psi_c^*\tilde\psi_s(x))\bigr)\\+a_-^*(\tilde\psi_s(x)^2+2\tilde\psi_s(x)\psi_c)\Bigr)e^{-\sigma x}\ ,
\end{multline}
and
\begin{multline}\label{eq:bminus}
b_-=\int_{-\infty}^\infty dx\,\bar\psi_{\xi}(x)^*\Bigl(2a_-\bigl(|\tilde\psi_s(x)|^2+2\re(\psi_c^*\tilde\psi_s(x))\bigr)\\+a_+^*(\tilde\psi_s(x)^2+2\tilde\psi_s(x)\psi_c)\Bigr)e^{-\sigma^*\! x}\ .
\end{multline}
The real form of \eqref{eq:osce} is
\begin{equation}\label{eq:osct}
\frac{d\xi}{dt}= Be^{-2\sigma_r\xi}\sin(2\sigma_i\xi-\beta)\ ,
\end{equation}
 using the polar representation $b_--b_+^*=Be^{-i\beta}$.
\begin{figure}[t]
\includegraphics[width=7.6cm]{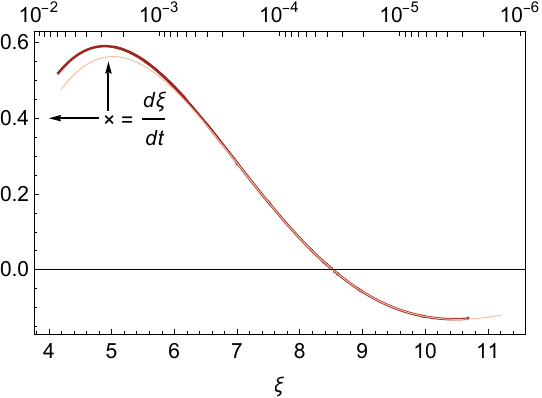}
\\[5mm]\includegraphics[width=7.35cm]{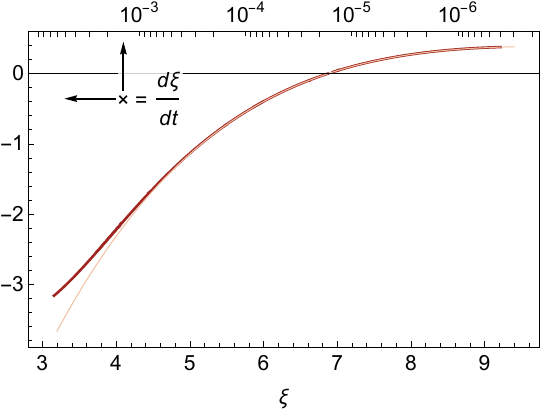}
\caption{\label{fig:osc} Kerr soliton drift velocity $d\xi/dt$ in waveforms with two oscillatory-tails solitons centered at $\xi$ and $-\xi$, for normalized cavity parameters $\delta=0.25$, $h_r=0.07$ (top) and $\delta=0.1$, $h_r=0.08$ (bottom). The significance of the thick dark curve and the thin light curve is the same as in figure \ref{fig:monotone}. The drift velocity oscillates as a function of the separation inside an exponentially decreasing envelope, with zero-velocity fixed points separating intervals of attraction and repulsion. The fixed point in the top panel is attracting, representing a stable pair molecule steady state, while the one in the right panel is repelling, separating initial soliton displacements that lead to an eventual collision, from ones where they drift apart.
Note that in order to increase the dynamic range, the drift velocity curves are multiplied by exponentially growing compensation factors in both panels, so that velocity values should be read from the graph by multiplying the left and top ordinates as indicated by the arrows.}
\end{figure}

Hence, the effective interaction between oscillatory-tail solitons can be either attractive or repulsive, depending on the separation, and the domains of attraction and repulsion are interlaced and separated by fixed points. In the asymptotic regime the fixed points are evenly spaced, and fixed points at $\xi=(\beta+n\pi)/(2\sigma_i)$ with $n$ even or odd are unstable or stable, respectively. 

Figure \ref{fig:osc} shows two examples of the soliton drift velocity in a pair waveform with oscillatory tails, comparing the asymptotic theory of \eqref{eq:osct} with calculations based on numerical solutions of the wave equation \eqref{eq:lle}. In one of the examples the solitons converge toward a stable stationary-molecule steady state, and in the second example the leading  fixed point of the effective dynamics is unstable.

We note that in the red-detuned regime $\sigma_r$ is comparable or larger than $\sigma_i$, so that even though the effective interaction has an infinite number of stable stationary pair molecule steady states, the exponential falloff of the interaction implies that the time needed to form high-order states is orders of magnitude longer than the duration of typical experiments, and since high-order states are very weakly bound, we anticipate that they are fragile to the effects external perturbations and noise.

\section{Conclusions}\label{sec:conclusions}
Recent experiments in pumped Kerr resonators revealed rich dynamics and interactions between bright optical solitons in multisoliton waveforms, while the theoretical investigation of this phenomenon has to date been limited. Here we study soliton pair interactions using the tail-overlap projection method, obtaining for the first time asymptotically exact expressions for the drift velocity of Kerr solitons in the large-separation limit. Since the solitons are exponentially localized, the interaction strength falls off exponentially with increasing separation, and, depending on whether the soliton tails are decreasing monotonically or are oscillatory, the interaction is either purely repulsive or sinusoidally modulated, switching periodically between attraction and repulsion. We found no instances of pure soliton attraction, although we have not been able to rule out this possibility.

The theory is facilitated by the large separation between the fast dynamical time scale on which the soliton shape evolves and the slow time scale of soliton drift. In microresonator experiments the dynamical time scale is of order nanosecond, so that drift velocities as slow as $10^{-9}$ can be observed. For red-detuned solitons, pair molecules can form during evolution times of $10^{10}$ or less if a stable fixed point of the effective dynamics is favorably placed, as in the top panel of figure \ref{fig:osc}; on the other hand, time scales for binding of larger-separation molecules are much too long for experimental realization, and there are cavity parameter values, like those of the bottom panel of figure \ref{fig:osc}, where no bound steady state is reachable. In physics applications, when the soliton separation is very large, effects that are not captured by the present theory become important, including fluctuations generated by noise and imperfections, and oscillatory pedestals associated with high-order dispersion.

Even though the effective interaction was derived here only for two-soliton waveforms, it is arguably adapted to multisoliton waveforms by observing that the rapid decay of interaction strength implies that in waveforms with more than two solitons the drift speed is dominated by the overlap of each soliton with its nearest neighbors, so that it is given by the two-body interactions to leading order in the overlap parameter $\varepsilon$. This argument explains the experimental and numerical observations of uniformly spaced multisoliton molecules and soliton crystals.
\bibliography{solipair}
\end{document}